\documentclass[aps,prl,reprint,superscriptaddress]{revtex4-1}

\usepackage[utf8]{inputenc}

\newcommand{\ghz}{~\mathrm{GHz}}
\newcommand{\mhz}{~\mathrm{MHz}}

\usepackage{ifpdf}

\ifpdf
\usepackage[pdftex]{graphicx}
\else
\usepackage{graphicx}
\fi

\begin{document}

\title{Fast Reset and Suppressing Spontaneous Emission of a Superconducting Qubit}

\author{M. D. Reed}
\affiliation{Departments of Physics and Applied Physics, Yale University, New Haven, Connecticut 06520, USA}
\author{B. R. Johnson}
\affiliation{Departments of Physics and Applied Physics, Yale University, New Haven, Connecticut 06520, USA}
\author{A. A. Houck}
\affiliation{Departments of Physics and Applied Physics, Yale University, New Haven, Connecticut 06520, USA}
\affiliation{Department of Electrical Engineering, Princeton University, Princeton, New Jersey 08544, USA}
\author{L. DiCarlo}
\affiliation{Departments of Physics and Applied Physics, Yale University, New Haven, Connecticut 06520, USA}
\author{J. M. Chow}
\affiliation{Departments of Physics and Applied Physics, Yale University, New Haven, Connecticut 06520, USA}
\author{D. I. Schuster}
\affiliation{Departments of Physics and Applied Physics, Yale University, New Haven, Connecticut 06520, USA}
\author{L. Frunzio}
\affiliation{Departments of Physics and Applied Physics, Yale University, New Haven, Connecticut 06520, USA}
\author{R. J. Schoelkopf}
\affiliation{Departments of Physics and Applied Physics, Yale University, New Haven, Connecticut 06520, USA}

\date{\today}

\ifpdf
\DeclareGraphicsExtensions{.pdf, .jpg, .tif}
\else
\DeclareGraphicsExtensions{.eps, .jpg}
\fi

\begin{abstract}
Spontaneous emission through a coupled cavity can be a significant decay channel for qubits in circuit quantum electrdynamics.  We present a circuit design that effectively eliminates spontaneous emission due to the Purcell effect while maintaining strong coupling to a low $Q$ cavity.  Excellent agreement over a wide range in frequency is found between measured qubit relaxation times and the predictions of a circuit model.  Using fast (nanosecond time-scale) flux biasing of the qubit, we demonstrate in-situ control of qubit lifetime over a factor of 50.  We realize qubit reset with 99.9\% fidelity in 120~ns.
\end{abstract}

\maketitle

In circuit quantum electrodynamics (cQED), engineered artificial atoms used as quantum bits (qubits) interact strongly with the electromagnetic modes of a transmission-line microwave cavity \cite{Wallraff2004}.  The large qubit-photon coupling affords capabilities such as coherent interactions of qubit and photon states \cite{Sillanpaa2007}, large coupling between spatially separated qubits mediated by the cavity bus \cite{Sillanpaa2007,Majer2007}, and non-destructive joint qubit readout \cite{Filipp2009,Chow2010}.  However, this strong coupling can also cause undesirable shortening of qubit lifetime ($T_{1}$) due to radiative decay through the cavity \cite{Houck2008}.  This effect, first described by E.~M. Purcell, describes a quantized system coupled to a resonant circuit or cavity \cite{Purcell1946}.  Depending on the detuning of the system transition frequency from the cavity resonance frequency, the rate of decay of the quantum system can be strongly enhanced \cite{Purcell1946,Goy1983} or suppressed \cite{Kleppner1981,Hulet1985,Jhe1987} compared with the decay rate to the electromagnetic continuum.  In cQED, qubits are generally sufficiently detuned to have suppressed relaxation, but $T_{1}$ can still be limited by decay through the cavity.  As qubit lifetime is of paramount importance in quantum computing \cite{Chow2009}, a means of further inhibiting radiative decay is desirable.

\begin{figure}
\centering
\includegraphics[scale=1]{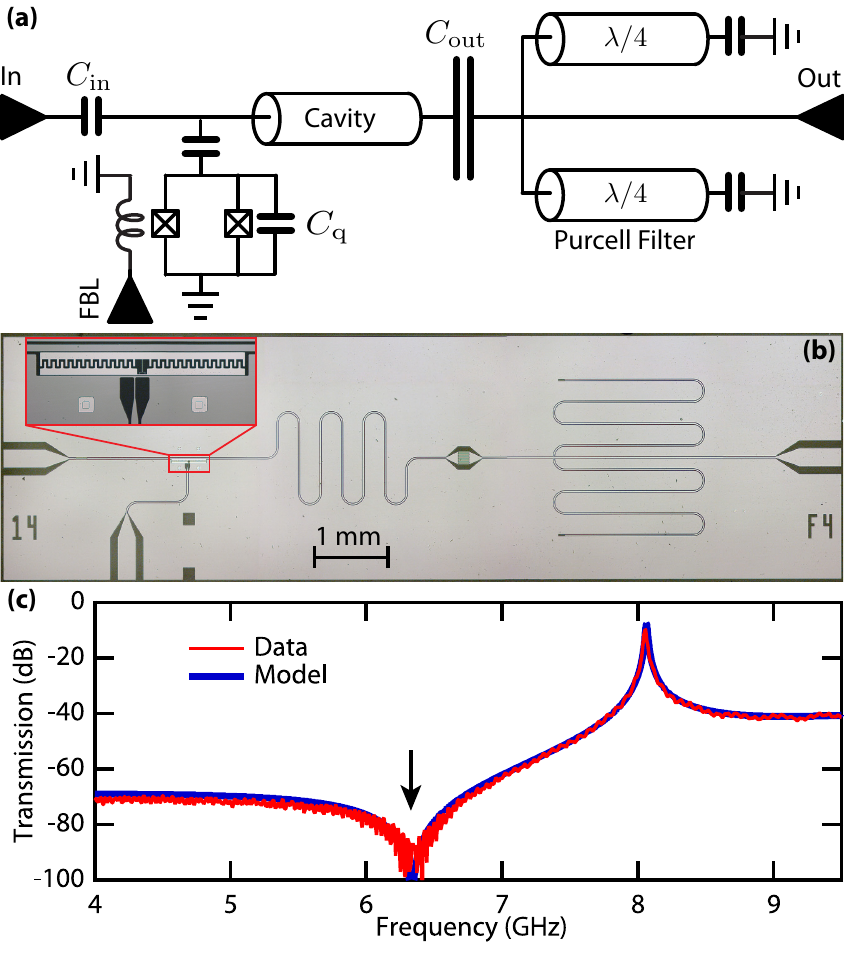}
\caption{Design, realization, and diagnostic transmission data of the Purcell filter.  (a) Circuit model of the Purcell-filtered cavity design.  The Purcell filter, implemented with twin $\lambda/4$ open-circuited transmission-line stubs, inhibits decay through $C_{\mathrm{out}}$ near its resonance $\omega_{\mathrm{f}}$.  (b) Optical micrograph of the device with inset zoom on transmon qubit.  Note the correspondence of the circuit elements directly above in (a).  (c) Cavity transmission measured at 4.2~K and comparison to the circuit-model prediction.  The Purcell filter shorts out the 50~$\Omega$ output environment at $\omega_{\mathrm{f}}$, yielding a 30~dB drop in transmission (arrow).  A circuit model involving only the parameters $C_{\mathrm{in}}$, $C_{\mathrm{out}}$, $\omega_{\mathrm{c}}$, and $\omega_{\mathrm{f}}$ shows excellent correspondence.}
{\label{fig:circuit}}
\end{figure}

The Purcell decay rate can be significantly reduced \cite{Houck2008} by increasing either the cavity quality factor $Q$ or the detuning between the qubit ($\omega_{\mathrm{q}}$) and cavity ($\omega_{\mathrm{c}}$) frequencies, $\Delta=\omega_{\mathrm{q}}-\omega_{\mathrm{c}}$, but these solutions have unwelcome implications of their own.  For example, reducing the cavity decay rate $\kappa = \omega_{\mathrm{c}} / Q$ can diminish qubit readout fidelity \cite{Gambetta2007} because fewer signal photons are collected in a qubit lifetime.  A large $\kappa$ is also beneficial for resetting a qubit to its ground state by bringing it near to the cavity resonance and exploiting the Purcell-enhanced decay rate.  Increasing $\Delta$ similarly has adverse effects on  readout fidelity and applications that exploit large state-dependent frequency shifts \cite{Schuster2007, DiCarlo2009, Chow2010}.  A better solution would improve qubit $T_{1}$ independent of the cavity $Q$, leaving its optimization up to other experimental concerns.

In this Letter, we introduce a design element for cQED termed the ``Purcell filter'', which protects a qubit from spontaneous emission while maintaining strong coupling to a low-$Q$ cavity.  We demonstrate an improvement of qubit $T_{1}$ by up to a factor of 50 compared to predicted values for an unfiltered device with the same $\kappa/2\pi\approx 20\mhz$.  Combining the large dynamic range of almost two orders of magnitude in $T_{1}$ with fast flux control, we then demonstrate fast qubit reset to 99\% (99.9\%) fidelity in 80~ns (120~ns).

The filter works by exploiting the fact that the qubit and cavity are typically far detuned.  We can therefore modify the qubit's electromagnetic environment (e.g. the density of photon states at $\omega_{\mathrm{q}}$) without, in principle, affecting the cavity $Q$ or resonant transmission.  The relationship between qubit $T_{1}$ due to spontaneous emission and admittance $Y$ of the coupled environment is  \begin{equation}T^{\mathrm{Purcell}}_{1}=\frac{C_{q}}{\mathrm{Re}[Y(\omega_{\mathrm{q}})]},\end{equation} where $C_{\mathrm{q}}$ is the qubit capacitance [Fig.~\ref{fig:circuit}(a)] \cite{Esteve1986,Neeley2008}. Previous work \cite{Houck2008} has demonstrated that Eq.~(1) accurately models the observed $T^{\mathrm{Purcell}}_{1}$ when all modes of the cavity are taken into account in the calculation of $Y$.  As the relationship holds for any admittance, this decay rate can be controlled by adjusting $Y$ with conventional microwave engineering techniques.  In particular, by manipulating $Y$ to be purely reactive (imaginary-valued) at $\omega_{\mathrm{q}}$, $T^{\mathrm{Purcell}}_{1}$ diverges and the Purcell decay channel is turned off.  This solution decouples the choice of cavity $Q$ from the Purcell decay rate as desired, and, as we will see,  has the advantage of using only conventional circuit elements placed in an experimentally convenient location.

\begin{figure}
\centering
\includegraphics[scale=1]{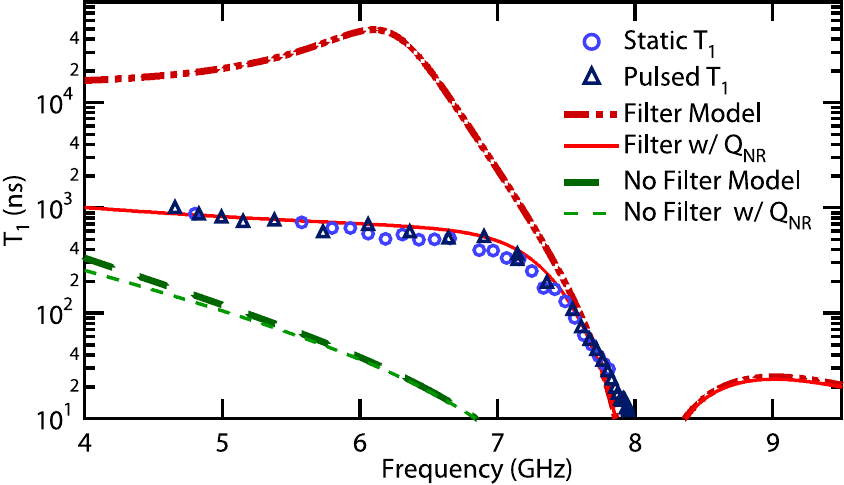}
\caption{Qubit $T_{1}$ as a function of frequency measured with two methods, and comparison to various models. The first method is a static measurement (circles): the qubit is excited and measured after a wait time $\tau$. The second (triangles) is a dynamic measurement: the qubit frequency is tuned with a fast flux pulse to an interrogation frequency, excited, and allowed to decay for $\tau$, and then returned to its operating frequency of $5.16\ghz$ and measured.  This method allows for accurate measurement even when $T_{1}$ is extremely short.  Measurements using the two methods show near perfect overlap. The top dashed curve is the predicted $T^{\mathrm{Purcell}}_{1}$, while solid curve includes also non-radiative internal loss with best-fit $Q_{\mathrm{NR}}=2\pi f T^{\mathrm{NR}}_{1}\approx27,000$.  The two lower curves correspond to an unfiltered device with the same $C_{\mathrm{in}}$, $C_{\mathrm{out}}$, and $\omega_{\mathrm{c}}$, with and without the internal loss. In this case, the Purcell filter gives a $T_{1}$ improvement by up to a factor of $\sim$50 ($6.7\ghz$).}
{\label{fig:lifetime}}
\end{figure}

We implement the Purcell filter with a transmission-line stub terminated in an open circuit placed outside the output capacitor $C_{\mathrm{out}}$ [Fig.~\ref{fig:circuit}(a)].  The length of this stub is set such that it acts as a $\lambda/4$ impedance transformer to short out the $50~\Omega$ environment at its resonance frequency $\omega_{\mathrm{f}}$.  We choose $C_{\mathrm{out}}$ to be much larger than the input capacitor, $C_{\mathrm{in}}\approx~C_{\mathrm{out}}/15$, to ensure that the qubit would be overwhelmingly likely to decay through $C_{\mathrm{out}}$.  The Purcell filter eliminates decay through this channel, leaving only the negligible decay rate through $C_{\mathrm{in}}$.  The combined total capacitance $C_{\mathrm{tot}}\approx80~\mathrm{fF}$ results in a small cavity $Q$.  We use two identical stubs above and below the major axis of the chip [Fig.~\ref{fig:circuit}(b)] to keep the design symmetric in an effort to suppress any undesired on-chip modes.  The cavity resonates at $\omega_{\mathrm{c}}/2\pi=8.04\ghz$, the filter at $\omega_{\mathrm{f}}/2\pi=6.33\ghz$, and a  flux bias line (FBL) is used to address a single transmon qubit \cite{Schreier2008} with a maximum frequency of $9.8\ghz$, a charging energy $E_{\mathrm{C}}/2\pi$ of $350\mhz$, and a resonator coupling strength $g/2\pi$ of $270\mhz$.  Transmission through the cavity measured at 4.2~K was compared with our model to validate the microwave characteristics of the device [Fig.~\ref{fig:circuit}(c)].  There is a dip corresponding to inhibited decay through $C_{\mathrm{out}}$ at $\omega_{\mathrm{f}}$.  The predicted and measured curves are also qualitatively similar, lending credence to the circuit model.  This method provided a convenient validation before cooling the device to 25~mK in a helium dilution refrigerator.

We measured the qubit $T_{1}$ as function of frequency and found it to be in excellent agreement with expectations.  $T_{1}$ is well modeled by the sum of the Purcell rate predicted by our filtered circuit model and a non-radiative internal loss $Q_{\mathrm{NR}}\approx27,000$ (Fig.~\ref{fig:lifetime}).  The source of this loss is a topic of current research, though some candidates are surface two level systems \cite{Shnirman2005,OConnell2008}, dielectric loss of the tunnel barrier oxide \cite{Martinis2005} or corundum substrate, and non-equilibrium quasiparticles \cite{Martinis2009}.   This model contains only the fit parameter $Q_{\mathrm{NR}}$ combined with the independently measured values of $g$, $E_{\mathrm{C}}$, $\omega_{\mathrm{c}}$, $\omega_{\mathrm{f}}$, $C_{\mathrm{in}}$, and $C_{\mathrm{out}}$.  An improvement to $T_{1}$ due to the Purcell filter was found to be as much as a factor of 50 at $6.7\ghz$ by comparison to an unfiltered circuit model with the same parameters.  This would be much greater in the absence of $Q_{\mathrm{NR}}$.  The device also exhibits a large dynamic range in $T_{1}$: about a factor of 80 between the longest and shortest times measured.

This range in $T_{1}$ can be a challenge to quantify because measurements made at small detunings, where $T_{1}$ is a few tens of nanoseconds, have a very low SNR.  This issue was avoided through the use of fast flux control \cite{DiCarlo2009}.  For measurements at small $\Delta$, the qubit is pulsed to the detuning under scrutiny, excited and allowed to decay, then pulsed to $5.16\ghz$ where measurement fidelity is higher, and interrogated.  In the cases where the qubit is nearly in resonance with the cavity, the $T_{1}$ is actually so short that it constitutes an interesting resource.

\begin{figure}
\centering
\includegraphics[scale=1]{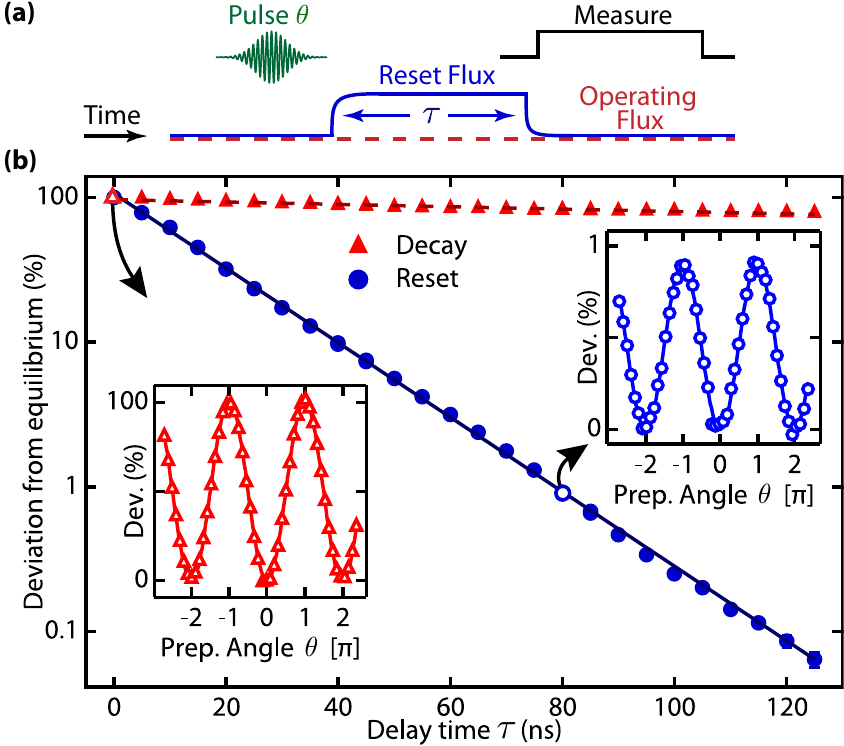}
\caption{Fast qubit reset.  (a) Schematic of a pulse sequence used to realize a qubit reset and characterize its performance.  The fidelity of reset was quantified using a modified Rabi oscillation scheme.  The qubit is first rotated around the $x$-axis by an angle $\theta$ at the operating frequency of $5.16\ghz$ and then pulsed into near resonance with the cavity (solid line) or left at the operating frequency (dashed line) for a time $\tau$.  The state of the qubit is measured as a function of $\theta$ and $\tau$ after being pulsed back to $5.16\ghz$.  
(b) The Rabi-oscillation amplitude as a function of $\tau$, normalized to the amplitude for $\tau$=0.  This ratio gives the deviation of the qubit state from equilibrium.  Curves are fit to exponentials with decay constants of 16.9~$\pm$~0.1~ns and 540~$\pm$~20~ns respectively. Insets: Measured Rabi oscillations for $\tau$=0 (lower left) and $\tau$=80~ns (top right).  The vertical scales differ by a factor of 100.}
{\label{fig:reset}}
\end{figure}

The ability to reset, or quickly cool a qubit to its equilibrium state on demand, is an important capability with a diverse set of applications.  Using a qubit to make repeated measurements of a coupled system, for example, requires resetting the qubit between interrogations \cite{Johnson2010}.  Similarly, experiment repetition rates can be greatly enhanced when they are otherwise limited by $T_{1}$.  Fast reset is also vital for measurement-free quantum error correction \cite{Mermin2007}.  In this scheme, an error syndrome is encoded in two ancilla qubits and conditionally corrected using a three qubit gate.  The ancillas, which now hold the entropy associated with the error, are then reset and reused. The Purcell filter is an ideal element with which to demonstrate reset as it allows for a relatively short reset time through the use of a low-$Q$ cavity without limiting $T_{1}$ at the operating frequency.

The efficacy of reset in this device is readily quantified using a modified Rabi oscillation sequence, described in Fig.~\ref{fig:reset}(a).  Each experiment measures the degree to which the qubit is out of equilibrium after some reset time $\tau$; the protocol is insensitive to any equilibrium thermal population of the qubit.  The non-equilibrium population is found to exhibit pure exponential decay over three orders of magnitude.  The qubit can be reset to 99.9\% in 120~ns or any other fidelity depending on $\tau$.  The sequence is also performed with the qubit remaining in the operating frequency during the delay to demonstrate the large dynamic range in $T_{1}$ available in this system.  In the case of multi-qubit devices, it is possible that this reset process would affect other qubits coupled to the same bus, but this issue could be avoided by using separate coupling and reset cavities.  

The Purcell filter is an important design element for cQED which allows for the use of low-$Q$ cavities without adversely affecting qubit $T_{1}$.  This ability is well-suited for in-situ qubit reset, a prerequisite for measurement-free quantum error correction and other applications.  We have observed high fidelity qubit measurements using both linear dispersive \cite{Wallraff2004} and Jaynes-Cummings readouts \cite{Reed2010} in this device, and a quantitative comparison with an unfiltered device remains for future work.

We acknowledge helpful discussions with M.~H.~Devoret and L.~S.~Bishop.  This research was supported by LPS/NSA under ARO Contract No. W911NF-05-1-0365, and by the NSF under Grants No. DMR-0653377 and No. DMR-0603369.  Additional support provided by CNR-Istituto di Cibernetica, Pozzuoli, Italy (LF).


\begin{thebibliography}{10}%
\makeatletter
\providecommand \@ifxundefined [1]{%
 \ifx #1\undefined \expandafter \@firstoftwo
 \else \expandafter \@secondoftwo
\fi
}%
\providecommand \@ifnum [1]{%
 \ifnum #1\expandafter \@firstoftwo
 \else \expandafter \@secondoftwo
\fi
}%
\providecommand \enquote [1]{``#1''}%
\providecommand \bibnamefont  [1]{#1}%
\providecommand \bibfnamefont [1]{#1}%
\providecommand \citenamefont [1]{#1}%
\providecommand\href[0]{\@sanitize\@href}%
\providecommand\@href[1]{\endgroup\@@startlink{#1}\endgroup\@@href}%
\providecommand\@@href[1]{#1\@@endlink}%
\providecommand \@sanitize [0]{\begingroup\catcode`\&12\catcode`\#12\relax}%
\@ifxundefined \pdfoutput {\@firstoftwo}{%
 \@ifnum{\z@=\pdfoutput}{\@firstoftwo}{\@secondoftwo}%
}{%
 \providecommand\@@startlink[1]{\leavevmode}%
 \providecommand\@@endlink[0]{}%
}{%
 \providecommand\@@startlink[1]{%
  \leavevmode
  \pdfstartlink
   attr{/Border[0 0 1 ]/H/I/C[0 1 1]}%
   user{/Subtype/Link/A<</Type/Action/S/URI/URI(#1)>>}%
  \relax
 }%
 \providecommand\@@endlink[0]{\pdfendlink}%
}%
\providecommand \url  [0]{\begingroup\@sanitize \@url }%
\providecommand \@url [1]{\endgroup\@href {#1}{\urlprefix}}%
\providecommand \urlprefix [0]{URL }%
\providecommand \Eprint[0]{\href }%
\@ifxundefined \urlstyle {%
  \providecommand \doi [1]{doi:\discretionary{}{}{}#1}%
}{%
  \providecommand \doi [0]{doi:\discretionary{}{}{}\begingroup
  \urlstyle{rm}\Url }%
}%
\providecommand \doibase [0]{http://dx.doi.org/}%
\providecommand \Doi[1]{\href{\doibase#1}}%
\providecommand \bibAnnote [3]{%
  \BibitemShut{#1}%
  \begin{quotation}\noindent
    \textsc{Key:}\ #2\\\textsc{Annotation:}\ #3%
  \end{quotation}%
}%
\providecommand \bibAnnoteFile [2]{%
  \IfFileExists{#2}{\bibAnnote {#1} {#2} {\input{#2}}}{}%
}%
\providecommand \typeout [0]{\immediate \write \m@ne }%
\providecommand \selectlanguage [0]{\@gobble}%
\providecommand \bibinfo [0]{\@secondoftwo}%
\providecommand \bibfield [0]{\@secondoftwo}%
\providecommand \translation [1]{[#1]}%
\providecommand \BibitemOpen[0]{}%
\providecommand \bibitemStop [0]{}%
\providecommand \bibitemNoStop [0]{.\EOS\space}%
\providecommand \EOS [0]{\spacefactor3000\relax}%
\providecommand \BibitemShut [1]{\csname bibitem#1\endcsname}%
\bibitem{Wallraff2004}%
  \BibitemOpen
  \bibfield{author}{%
  \bibinfo {author} {\bibfnamefont{A.}~\bibnamefont{Wallraff}}, \bibinfo
  {author} {\bibfnamefont{D.~I.}\ \bibnamefont{Schuster}}, \bibinfo {author}
  {\bibfnamefont{A.}~\bibnamefont{Blais}}, \bibinfo {author}
  {\bibfnamefont{L.}~\bibnamefont{Frunzio}}, \bibinfo {author}
  {\bibfnamefont{R.~S.}\ \bibnamefont{Huang}}, \bibinfo {author}
  {\bibfnamefont{J.}~\bibnamefont{Majer}}, \bibinfo {author}
  {\bibfnamefont{S.}~\bibnamefont{Kumar}}, \bibinfo {author}
  {\bibfnamefont{S.~M.}\ \bibnamefont{Girvin}},\ and\ \bibinfo {author}
  {\bibfnamefont{R.~J.}\ \bibnamefont{Schoelkopf}},\ }%
  \bibfield{journal}{%
  \bibinfo {journal} {Nature}\ }%
  \textbf{\bibinfo {volume} {431}},\ \bibinfo {pages} {162} (\bibinfo {year}
  {2004})%
  \bibAnnoteFile{NoStop}{Wallraff2004}%
\bibitem{Sillanpaa2007}%
  \BibitemOpen
  \bibfield{author}{%
  \bibinfo {author} {\bibfnamefont{M.~A.}\ \bibnamefont{Sillanp{\"a}{\"a}}},
  \bibinfo {author} {\bibfnamefont{J.~I.}\ \bibnamefont{Park}},\ and\ \bibinfo
  {author} {\bibfnamefont{R.~W.}\ \bibnamefont{Simmonds}},\ }%
  \bibfield{journal}{%
  \bibinfo {journal} {Nature}\ }%
  \textbf{\bibinfo {volume} {449}},\ \bibinfo {pages} {438} (\bibinfo {year}
  {2007})%
  \bibAnnoteFile{NoStop}{Sillanpaa2007}%
\bibitem{Majer2007}%
  \BibitemOpen
  \bibfield{author}{%
  \bibinfo {author} {\bibfnamefont{J.}~\bibnamefont{Majer}}, \bibinfo {author}
  {\bibfnamefont{J.~M.}\ \bibnamefont{Chow}}, \bibinfo {author}
  {\bibfnamefont{J.~M.}\ \bibnamefont{Gambetta}}, \bibinfo {author}
  {\bibfnamefont{J.}~\bibnamefont{Koch}}, \bibinfo {author}
  {\bibfnamefont{B.~R.}\ \bibnamefont{Johnson}}, \bibinfo {author}
  {\bibfnamefont{J.~A.}\ \bibnamefont{Schreier}}, \bibinfo {author}
  {\bibfnamefont{L.}~\bibnamefont{Frunzio}}, \bibinfo {author}
  {\bibfnamefont{D.~I.}\ \bibnamefont{Schuster}}, \bibinfo {author}
  {\bibfnamefont{A.~A.}\ \bibnamefont{Houck}}, \bibinfo {author}
  {\bibfnamefont{A.}~\bibnamefont{Wallraff}}, \bibinfo {author}
  {\bibfnamefont{A.}~\bibnamefont{Blais}}, \bibinfo {author}
  {\bibfnamefont{M.~H.}\ \bibnamefont{Devoret}}, \bibinfo {author}
  {\bibfnamefont{S.~M.}\ \bibnamefont{Girvin}},\ and\ \bibinfo {author}
  {\bibfnamefont{R.~J.}\ \bibnamefont{Schoelkopf}},\ }%
  \bibfield{journal}{%
  \bibinfo {journal} {Nature}\ }%
  \textbf{\bibinfo {volume} {449}},\ \bibinfo {pages} {443} (\bibinfo {year}
  {2007})%
  \bibAnnoteFile{NoStop}{Majer2007}%
\bibitem{Filipp2009}%
  \BibitemOpen
  \bibfield{author}{%
  \bibinfo {author} {\bibfnamefont{S.}~\bibnamefont{Filipp}}, \bibinfo {author}
  {\bibfnamefont{P.}~\bibnamefont{Maurer}}, \bibinfo {author}
  {\bibfnamefont{P.~J.}\ \bibnamefont{Leek}}, \bibinfo {author}
  {\bibfnamefont{M.}~\bibnamefont{Baur}}, \bibinfo {author}
  {\bibfnamefont{R.}~\bibnamefont{Bianchetti}}, \bibinfo {author}
  {\bibfnamefont{J.~M.}\ \bibnamefont{Fink}}, \bibinfo {author}
  {\bibfnamefont{M.}~\bibnamefont{G{\"o}ppl}}, \bibinfo {author}
  {\bibfnamefont{L.}~\bibnamefont{Steffen}}, \bibinfo {author}
  {\bibfnamefont{J.~M.}\ \bibnamefont{Gambetta}}, \bibinfo {author}
  {\bibfnamefont{A.}~\bibnamefont{Blais}},\ and\ \bibinfo {author}
  {\bibfnamefont{A.}~\bibnamefont{Wallraff}},\ }%
  \bibfield{journal}{%
  \bibinfo {journal} {Phys. Rev. Lett.}\ }%
  \textbf{\bibinfo {volume} {102}},\ \bibinfo {pages} {200402} (\bibinfo {year}
  {2009})%
  \bibAnnoteFile{NoStop}{Filipp2009}%
\bibitem{Chow2010}%
  \BibitemOpen
  \bibfield{author}{%
  \bibinfo {author} {\bibfnamefont{J.~M.}\ \bibnamefont{Chow}}, \bibinfo
  {author} {\bibfnamefont{L.}~\bibnamefont{DiCarlo}}, \bibinfo {author}
  {\bibfnamefont{J.~M.}\ \bibnamefont{Gambetta}}, \bibinfo {author}
  {\bibfnamefont{A.}~\bibnamefont{Nunnenkamp}}, \bibinfo {author}
  {\bibfnamefont{L.~S.}\ \bibnamefont{Bishop}}, \bibinfo {author}
  {\bibfnamefont{L.}~\bibnamefont{Frunzio}}, \bibinfo {author}
  {\bibfnamefont{M.~H.}\ \bibnamefont{Devoret}}, \bibinfo {author}
  {\bibfnamefont{S.~M.}\ \bibnamefont{Girvin}},\ and\ \bibinfo {author}
  {\bibfnamefont{R.~J.}\ \bibnamefont{Schoelkopf}},\ }%
  \bibinfo {journal} {arXiv:0908.1955}%
  \bibAnnoteFile{NoStop}{Chow2010}%
\bibitem{Houck2008}%
  \BibitemOpen
\bibfield{journal}{%
    }%
  \bibfield{author}{%
  \bibinfo {author} {\bibfnamefont{A.~A.}\ \bibnamefont{Houck}}, \bibinfo
  {author} {\bibfnamefont{J.~A.}\ \bibnamefont{Schreier}}, \bibinfo {author}
  {\bibfnamefont{B.~R.}\ \bibnamefont{Johnson}}, \bibinfo {author}
  {\bibfnamefont{J.~M.}\ \bibnamefont{Chow}}, \bibinfo {author}
  {\bibfnamefont{J.}~\bibnamefont{Koch}}, \bibinfo {author}
  {\bibfnamefont{J.~M.}\ \bibnamefont{Gambetta}}, \bibinfo {author}
  {\bibfnamefont{D.~I.}\ \bibnamefont{Schuster}}, \bibinfo {author}
  {\bibfnamefont{L.}~\bibnamefont{Frunzio}}, \bibinfo {author}
  {\bibfnamefont{M.~H.}\ \bibnamefont{Devoret}}, \bibinfo {author}
  {\bibfnamefont{S.~M.}\ \bibnamefont{Girvin}},\ and\ \bibinfo {author}
  {\bibfnamefont{R.~J.}\ \bibnamefont{Schoelkopf}},\ }%
  \bibfield{journal}{%
  \bibinfo {journal} {Phys. Rev. Lett.}\ }%
  \textbf{\bibinfo {volume} {101}},\ \bibinfo {pages} {080502} (\bibinfo {year}
  {2008})%
  \bibAnnoteFile{NoStop}{Houck2008}%
\bibitem{Purcell1946}%
  \BibitemOpen
  \bibfield{author}{%
  \bibinfo {author} {\bibfnamefont{E.~M.}\ \bibnamefont{Purcell}},\ }%
  \bibfield{journal}{%
  \bibinfo {journal} {Phys. Rev.}\ }%
  \textbf{\bibinfo {volume} {69}},\ \bibinfo {pages} {681} (\bibinfo {year}
  {1946})%
  \bibAnnoteFile{NoStop}{Purcell1946}%
\bibitem{Goy1983}%
  \BibitemOpen
  \bibfield{author}{%
  \bibinfo {author} {\bibfnamefont{P.}~\bibnamefont{Goy}}, \bibinfo {author}
  {\bibfnamefont{J.~M.}\ \bibnamefont{Raimond}}, \bibinfo {author}
  {\bibfnamefont{M.}~\bibnamefont{Gross}},\ and\ \bibinfo {author}
  {\bibfnamefont{S.}~\bibnamefont{Haroche}},\ }%
  \bibfield{journal}{%
  \bibinfo {journal} {Phys. Rev. Lett.}\ }%
  \textbf{\bibinfo {volume} {50}},\ \bibinfo {pages} {1903} (\bibinfo {year}
  {1983})%
  \bibAnnoteFile{NoStop}{Goy1983}%
\bibitem{Kleppner1981}%
  \BibitemOpen
  \bibfield{author}{%
  \bibinfo {author} {\bibfnamefont{D.}~\bibnamefont{Kleppner}},\ }%
  \bibfield{journal}{%
  \bibinfo {journal} {Phys. Rev. Lett.}\ }%
  \textbf{\bibinfo {volume} {47}},\ \bibinfo {pages} {233} (\bibinfo {year}
  {1981})%
  \bibAnnoteFile{NoStop}{Kleppner1981}%
\bibitem{Hulet1985}%
  \BibitemOpen
  \bibfield{author}{%
  \bibinfo {author} {\bibfnamefont{R.~G.}\ \bibnamefont{Hulet}}, \bibinfo
  {author} {\bibfnamefont{E.~S.}\ \bibnamefont{Hilfer}},\ and\ \bibinfo
  {author} {\bibfnamefont{D.}~\bibnamefont{Kleppner}},\ }%
  \bibfield{journal}{%
  \bibinfo {journal} {Phys. Rev. Lett.}\ }%
  \textbf{\bibinfo {volume} {55}},\ \bibinfo {pages} {2137} (\bibinfo {year}
  {1985})%
  \bibAnnoteFile{NoStop}{Hulet1985}%
\bibitem{Jhe1987}%
  \BibitemOpen
  \bibfield{author}{%
  \bibinfo {author} {\bibfnamefont{W.}~\bibnamefont{Jhe}}, \bibinfo {author}
  {\bibfnamefont{A.}~\bibnamefont{Anderson}}, \bibinfo {author}
  {\bibfnamefont{E.~A.}\ \bibnamefont{Hinds}}, \bibinfo {author}
  {\bibfnamefont{D.}~\bibnamefont{Meschede}}, \bibinfo {author}
  {\bibfnamefont{L.}~\bibnamefont{Moi}},\ and\ \bibinfo {author}
  {\bibfnamefont{S.}~\bibnamefont{Haroche}},\ }%
  \bibfield{journal}{%
  \bibinfo {journal} {Phys. Rev. Lett.}\ }%
  \textbf{\bibinfo {volume} {58}},\ \bibinfo {pages} {666} (\bibinfo {year}
  {1987})%
  \bibAnnoteFile{NoStop}{Jhe1987}%
\bibitem{Chow2009}%
  \BibitemOpen
  \bibfield{author}{%
  \bibinfo {author} {\bibfnamefont{J.~M.}\ \bibnamefont{Chow}}, \bibinfo
  {author} {\bibfnamefont{J.~M.}\ \bibnamefont{Gambetta}}, \bibinfo {author}
  {\bibfnamefont{L.}~\bibnamefont{Tornberg}}, \bibinfo {author}
  {\bibfnamefont{J.}~\bibnamefont{Koch}}, \bibinfo {author}
  {\bibfnamefont{L.~S.}\ \bibnamefont{Bishop}}, \bibinfo {author}
  {\bibfnamefont{A.~A.}\ \bibnamefont{Houck}}, \bibinfo {author}
  {\bibfnamefont{B.~R.}\ \bibnamefont{Johnson}}, \bibinfo {author}
  {\bibfnamefont{L.}~\bibnamefont{Frunzio}}, \bibinfo {author}
  {\bibfnamefont{S.~M.}\ \bibnamefont{Girvin}},\ and\ \bibinfo {author}
  {\bibfnamefont{R.~J.}\ \bibnamefont{Schoelkopf}},\ }%
  \bibfield{journal}{%
  \bibinfo {journal} {Phys. Rev. Lett.}\ }%
  \textbf{\bibinfo {volume} {102}},\ \bibinfo {pages} {090502} (\bibinfo {year}
  {2009})%
  \bibAnnoteFile{NoStop}{Chow2009}%
\bibitem{Gambetta2007}%
  \BibitemOpen
  \bibfield{author}{%
  \bibinfo {author} {\bibfnamefont{J.}~\bibnamefont{Gambetta}}, \bibinfo
  {author} {\bibfnamefont{W.~A.}\ \bibnamefont{Braff}}, \bibinfo {author}
  {\bibfnamefont{A.}~\bibnamefont{Wallraff}}, \bibinfo {author}
  {\bibfnamefont{S.~M.}\ \bibnamefont{Girvin}},\ and\ \bibinfo {author}
  {\bibfnamefont{R.~J.}\ \bibnamefont{Schoelkopf}},\ }%
  \bibfield{journal}{%
  \bibinfo {journal} {Phys. Rev. A}\ }%
  \textbf{\bibinfo {volume} {76}},\ \bibinfo {pages} {012325} (\bibinfo {year}
  {2007})%
  \bibAnnoteFile{NoStop}{Gambetta2007}%
\bibitem{Schuster2007}%
  \BibitemOpen
  \bibfield{author}{%
  \bibinfo {author} {\bibfnamefont{D.~I.}\ \bibnamefont{Schuster}}, \bibinfo
  {author} {\bibfnamefont{A.~A.}\ \bibnamefont{Houck}}, \bibinfo {author}
  {\bibfnamefont{J.~A.}\ \bibnamefont{Schreier}}, \bibinfo {author}
  {\bibfnamefont{A.}~\bibnamefont{Wallraff}}, \bibinfo {author}
  {\bibfnamefont{J.~M.}\ \bibnamefont{Gambetta}}, \bibinfo {author}
  {\bibfnamefont{A.}~\bibnamefont{Blais}}, \bibinfo {author}
  {\bibfnamefont{L.}~\bibnamefont{Frunzio}}, \bibinfo {author}
  {\bibfnamefont{J.}~\bibnamefont{Majer}}, \bibinfo {author}
  {\bibfnamefont{B.}~\bibnamefont{Johnson}}, \bibinfo {author}
  {\bibfnamefont{M.~H.}\ \bibnamefont{Devoret}}, \bibinfo {author}
  {\bibfnamefont{S.~M.}\ \bibnamefont{Girvin}},\ and\ \bibinfo {author}
  {\bibfnamefont{R.~J.}\ \bibnamefont{Schoelkopf}},\ }%
  \bibfield{journal}{%
  \bibinfo {journal} {Nature}\ }%
  \textbf{\bibinfo {volume} {445}},\ \bibinfo {pages} {515} (\bibinfo {year}
  {2007})%
  \bibAnnoteFile{NoStop}{Schuster2007}%
\bibitem{DiCarlo2009}%
  \BibitemOpen
  \bibfield{author}{%
  \bibinfo {author} {\bibfnamefont{L.}~\bibnamefont{DiCarlo}}, \bibinfo
  {author} {\bibfnamefont{J.~M.}\ \bibnamefont{Chow}}, \bibinfo {author}
  {\bibfnamefont{J.~M.}\ \bibnamefont{Gambetta}}, \bibinfo {author}
  {\bibfnamefont{L.~S.}\ \bibnamefont{Bishop}}, \bibinfo {author}
  {\bibfnamefont{B.~R.}\ \bibnamefont{Johnson}}, \bibinfo {author}
  {\bibfnamefont{D.~I.}\ \bibnamefont{Schuster}}, \bibinfo {author}
  {\bibfnamefont{J.}~\bibnamefont{Majer}}, \bibinfo {author}
  {\bibfnamefont{A.}~\bibnamefont{Blais}}, \bibinfo {author}
  {\bibfnamefont{L.}~\bibnamefont{Frunzio}}, \bibinfo {author}
  {\bibfnamefont{S.~M.}\ \bibnamefont{Girvin}},\ and\ \bibinfo {author}
  {\bibfnamefont{R.~J.}\ \bibnamefont{Schoelkopf}},\ }%
  \bibfield{journal}{%
  \bibinfo {journal} {Nature}\ }%
  \textbf{\bibinfo {volume} {460}},\ \bibinfo {pages} {240} (\bibinfo {year}
  {2009})%
  \bibAnnoteFile{NoStop}{DiCarlo2009}%
\bibitem{Esteve1986}%
  \BibitemOpen
  \bibfield{author}{%
  \bibinfo {author} {\bibfnamefont{D.}~\bibnamefont{Esteve}}, \bibinfo {author}
  {\bibfnamefont{M.~H.}\ \bibnamefont{Devoret}},\ and\ \bibinfo {author}
  {\bibfnamefont{J.~M.}\ \bibnamefont{Martinis}},\ }%
  \bibfield{journal}{%
  \bibinfo {journal} {Phys. Rev. B}\ }%
  \textbf{\bibinfo {volume} {34}},\ \bibinfo {pages} {158} (\bibinfo {year}
  {1986})%
  \bibAnnoteFile{NoStop}{Esteve1986}%
\bibitem{Neeley2008}%
  \BibitemOpen
  \bibfield{author}{%
  \bibinfo {author} {\bibfnamefont{M.}~\bibnamefont{Neeley}}, \bibinfo {author}
  {\bibfnamefont{M.}~\bibnamefont{Ansmann}}, \bibinfo {author}
  {\bibfnamefont{R.~C.}\ \bibnamefont{Bialczak}}, \bibinfo {author}
  {\bibfnamefont{M.}~\bibnamefont{Hofheinz}}, \bibinfo {author}
  {\bibfnamefont{N.}~\bibnamefont{Katz}}, \bibinfo {author}
  {\bibfnamefont{E.}~\bibnamefont{Lucero}}, \bibinfo {author}
  {\bibfnamefont{A.}~\bibnamefont{O'Connell}}, \bibinfo {author}
  {\bibfnamefont{H.}~\bibnamefont{Wang}}, \bibinfo {author}
  {\bibfnamefont{A.~N.}\ \bibnamefont{Cleland}},\ and\ \bibinfo {author}
  {\bibfnamefont{J.~M.}\ \bibnamefont{Martinis}},\ }%
  \bibfield{journal}{%
  \bibinfo {journal} {Phys. Rev. B}\ }%
  \textbf{\bibinfo {volume} {77}},\ \bibinfo {pages} {180508} (\bibinfo {year}
  {2008})%
  \bibAnnoteFile{NoStop}{Neeley2008}%
\bibitem{Schreier2008}%
  \BibitemOpen
  \bibfield{author}{%
  \bibinfo {author} {\bibfnamefont{J.~A.}\ \bibnamefont{Schreier}}, \bibinfo
  {author} {\bibfnamefont{A.~A.}\ \bibnamefont{Houck}}, \bibinfo {author}
  {\bibfnamefont{J.}~\bibnamefont{Koch}}, \bibinfo {author}
  {\bibfnamefont{D.~I.}\ \bibnamefont{Schuster}}, \bibinfo {author}
  {\bibfnamefont{B.~R.}\ \bibnamefont{Johnson}}, \bibinfo {author}
  {\bibfnamefont{J.~M.}\ \bibnamefont{Chow}}, \bibinfo {author}
  {\bibfnamefont{J.~M.}\ \bibnamefont{Gambetta}}, \bibinfo {author}
  {\bibfnamefont{J.}~\bibnamefont{Majer}}, \bibinfo {author}
  {\bibfnamefont{L.}~\bibnamefont{Frunzio}}, \bibinfo {author}
  {\bibfnamefont{M.~H.}\ \bibnamefont{Devoret}}, \bibinfo {author}
  {\bibfnamefont{S.~M.}\ \bibnamefont{Girvin}},\ and\ \bibinfo {author}
  {\bibfnamefont{R.~J.}\ \bibnamefont{Schoelkopf}},\ }%
  \bibfield{journal}{%
  \bibinfo {journal} {Phys. Rev. B}\ }%
  \textbf{\bibinfo {volume} {77}},\ \bibinfo {pages} {180502} (\bibinfo {year}
  {2008})%
  \bibAnnoteFile{NoStop}{Schreier2008}%
\bibitem{Shnirman2005}%
  \BibitemOpen
  \bibfield{author}{%
  \bibinfo {author} {\bibfnamefont{A.}~\bibnamefont{Shnirman}}, \bibinfo
  {author} {\bibfnamefont{G.}~\bibnamefont{Sch{\"o}n}}, \bibinfo {author}
  {\bibfnamefont{I.}~\bibnamefont{Martin}},\ and\ \bibinfo {author}
  {\bibfnamefont{Y.}~\bibnamefont{Makhlin}},\ }%
  \bibfield{journal}{%
  \bibinfo {journal} {Phys. Rev. Lett.}\ }%
  \textbf{\bibinfo {volume} {94}},\ \bibinfo {pages} {127002} (\bibinfo {year}
  {2005})%
  \bibAnnoteFile{NoStop}{Shnirman2005}%
\bibitem{OConnell2008}%
  \BibitemOpen
  \bibfield{author}{%
  \bibinfo {author} {\bibfnamefont{A.~D.}\ \bibnamefont{O'Connell}}, \bibinfo
  {author} {\bibfnamefont{M.}~\bibnamefont{Ansmann}}, \bibinfo {author}
  {\bibfnamefont{R.~C.}\ \bibnamefont{Bialczak}}, \bibinfo {author}
  {\bibfnamefont{M.}~\bibnamefont{Hofheinz}}, \bibinfo {author}
  {\bibfnamefont{N.}~\bibnamefont{Katz}}, \bibinfo {author}
  {\bibfnamefont{E.}~\bibnamefont{Lucero}}, \bibinfo {author}
  {\bibfnamefont{C.}~\bibnamefont{McKenney}}, \bibinfo {author}
  {\bibfnamefont{M.}~\bibnamefont{Neeley}}, \bibinfo {author}
  {\bibfnamefont{H.}~\bibnamefont{Wang}}, \bibinfo {author}
  {\bibfnamefont{E.~M.}\ \bibnamefont{Weig}}, \bibinfo {author}
  {\bibfnamefont{A.~N.}\ \bibnamefont{Cleland}},\ and\ \bibinfo {author}
  {\bibfnamefont{J.~M.}\ \bibnamefont{Martinis}},\ }%
  \bibfield{journal}{%
  \bibinfo {journal} {Appl. Phys. Lett.}\ }%
  \textbf{\bibinfo {volume} {92}},\ \bibinfo {pages} {112903} (\bibinfo {year}
  {2008})%
  \bibAnnoteFile{NoStop}{OConnell2008}%
\bibitem{Martinis2005}%
  \BibitemOpen
  \bibfield{author}{%
  \bibinfo {author} {\bibfnamefont{J.~M.}\ \bibnamefont{Martinis}}, \bibinfo
  {author} {\bibfnamefont{K.~B.}\ \bibnamefont{Cooper}}, \bibinfo {author}
  {\bibfnamefont{R.}~\bibnamefont{McDermott}}, \bibinfo {author}
  {\bibfnamefont{M.}~\bibnamefont{Steffen}}, \bibinfo {author}
  {\bibfnamefont{M.}~\bibnamefont{Ansmann}}, \bibinfo {author}
  {\bibfnamefont{K.~D.}\ \bibnamefont{Osborn}}, \bibinfo {author}
  {\bibfnamefont{K.}~\bibnamefont{Cicak}}, \bibinfo {author}
  {\bibfnamefont{S.}~\bibnamefont{Oh}}, \bibinfo {author}
  {\bibfnamefont{D.~P.}\ \bibnamefont{Pappas}}, \bibinfo {author}
  {\bibfnamefont{R.~W.}\ \bibnamefont{Simmonds}},\ and\ \bibinfo {author}
  {\bibfnamefont{C.~C.}\ \bibnamefont{Yu}},\ }%
  \bibfield{journal}{%
  \bibinfo {journal} {Phys. Rev. Lett.}\ }%
  \textbf{\bibinfo {volume} {95}},\ \bibinfo {pages} {210503} (\bibinfo {year}
  {2005})%
  \bibAnnoteFile{NoStop}{Martinis2005}%
\bibitem{Martinis2009}%
  \BibitemOpen
  \bibfield{author}{%
  \bibinfo {author} {\bibfnamefont{J.~M.}\ \bibnamefont{Martinis}}, \bibinfo
  {author} {\bibfnamefont{M.}~\bibnamefont{Ansmann}},\ and\ \bibinfo {author}
  {\bibfnamefont{J.}~\bibnamefont{Aumentado}},\ }%
  \bibfield{journal}{%
  \bibinfo {journal} {Phys. Rev. Lett.}\ }%
  \textbf{\bibinfo {volume} {103}},\ \bibinfo {pages} {097002} (\bibinfo {year}
  {2009})%
  \bibAnnoteFile{NoStop}{Martinis2009}%
\bibitem{Johnson2010}%
  \BibitemOpen
  \bibfield{author}{%
  \bibinfo {author} {\bibfnamefont{B.~R.}\ \bibnamefont{Johnson}}, \bibinfo
  {author} {\bibfnamefont{M.~D.}\ \bibnamefont{Reed}}, \bibinfo {author}
  {\bibfnamefont{A.~A.}\ \bibnamefont{Houck}}, \bibinfo {author}
  {\bibfnamefont{D.~I.}\ \bibnamefont{Schuster}}, \bibinfo {author}
  {\bibfnamefont{L.~S.}\ \bibnamefont{Bishop}}, \bibinfo {author}
  {\bibfnamefont{E.}~\bibnamefont{Ginossar}}, \bibinfo {author}
  {\bibfnamefont{J.~M.}\ \bibnamefont{Gambetta}}, \bibinfo {author}
  {\bibfnamefont{L.}~\bibnamefont{DiCarlo}}, \bibinfo {author}
  {\bibfnamefont{L.}~\bibnamefont{Frunzio}}, \bibinfo {author}
  {\bibfnamefont{S.~M.}\ \bibnamefont{Girvin}},\ and\ \bibinfo {author}
  {\bibfnamefont{R.~J.}\ \bibnamefont{Schoelkopf}},\ }%
  \bibfield{journal}{%
  \bibinfo {journal} {arXiv:1003.2734}}%
   (\bibinfo {year} {2010})%
  \bibAnnoteFile{NoStop}{Johnson2010}%
\bibitem{Mermin2007}%
  \BibitemOpen
  \bibfield{author}{%
  \bibinfo {author} {\bibfnamefont{N.~D.}\ \bibnamefont{Mermin}},\ }%
  \emph{\bibinfo {title} {Quantum Computer Science}}\ (\bibinfo {publisher}
  {Cambridge Univeristy Press, New York},\ \bibinfo {year} {2007})%
  \bibAnnoteFile{NoStop}{Mermin2007}%
\bibitem{Reed2010}%
  \BibitemOpen
  \bibfield{author}{%
  \bibinfo {author} {\bibfnamefont{M.~D.}\ \bibnamefont{Reed}}, \bibinfo
  {author} {\bibfnamefont{L.}~\bibnamefont{DiCarlo}}, \bibinfo {author}
  {\bibfnamefont{B.~R.}\ \bibnamefont{Johnson}}, \bibinfo {author}
  {\bibfnamefont{L.}~\bibnamefont{Sun}}, \bibinfo {author}
  {\bibfnamefont{D.~I.}\ \bibnamefont{Schuster}}, \bibinfo {author}
  {\bibfnamefont{L.}~\bibnamefont{Frunzio}},\ and\ \bibinfo {author}
  {\bibfnamefont{R.~J.}\ \bibnamefont{Schoelkopf}},\ }%
  \bibfield{journal}{%
  \bibinfo {journal} {arXiv:1004.4323}}%
   (\bibinfo {year} {2010})%
  \bibAnnoteFile{NoStop}{Reed2010}%
\end{thebibliography}
%

\end{document}